\documentclass[11pt]{article}
\usepackage{moriond,epsfig}

\def\be{\begin{equation}}
\def\ee{\end{equation}}
\def\bea{\begin{eqnarray}}
\def\eea{\end{eqnarray}}

\begin{document}
\vspace*{4cm}
\title{MEASUREMENT OF GAUGE BOSON COUPLINGS AND\\
W SPIN DENSITY MATRIX AT LEP}

\author{ EVELYNE DELMEIRE }

\address{D\'epartement de Physique Nucl\'eaire et Corpusculaire,\\
Universit\'e de Gen\`eve, Switzerland}

\maketitle\abstracts{
During the LEP2 period the $\rm{e}^{+}\rm{e}^{-}$ collider increased
its center of mass energy from 161 GeV
to 209 GeV and a total integrated luminosity of approximately 700 
$\rm{pb}^{-1}$ was recorded per experiment.
Pairs of $\rm{W}$ bosons are produced and allow the study
of gauge boson couplings involving 
$\rm{W}$, $\rm{Z}$ and photon. The coupling
of the $\rm{W}$ boson to the neutral gauge bosons
have been measured to be
$g_{1}^{Z}= 0.998^{+.023}_{-.025}$,
$\kappa_{\gamma}= 0.943^{+.055}_{-.055}$,
and $\lambda_{\gamma}= -0.020^{+.024}_{-.024}$ and are in agreement
with the Standard Model expectation. 
%of $g_{1}^{Z}=1$,
%$\kappa_{\gamma}=1$,
%and $\lambda_{\gamma}=0$
Limits are set on $\rm{CP}$-violating couplings
by a Spin Density Matrix analysis of the $\rm{W}$ decay products.
No evidence has been found for couplings of three neutral
gauge bosons, parametrized by $f_{4,5}^{Z,\gamma}$
and $h_{1,2,3,4}^{Z,\gamma}$. 
Limits are derived on couplings of four
gauge bosons, parametrized by
$a_{0}^{Z,W}/\Lambda^{2}$, $a_{n}^{W}/\Lambda^{2}$
and $a_{c}^{Z,W}/\Lambda^{2}$
where $\Lambda$ represents the energy scale for new physics.}

\section{Gauge Boson Couplings}
At LEP2, data are collected at center of mass energies ranging from 161 GeV up to 
209 GeV. Massive $\rm{W}$ bosons are produced in pairs via 
$\rm{e}^{+}\rm{e}^{-}$ interactions
and gauge  
couplings involving $\rm{W}$, $\rm{Z}$ and photon are studied
by the four LEP experiments ALEPH, DELPHI, L3 and OPAL~\cite{LEPEWWG}.\\
\indent
The non-Abelian $SU(2)_{L}\otimes U(1)_{Y}$ gauge symmetry of the 
Standard Model~\cite{GWS}
predicts tree level interactions between three or four 
charged and neutral gauge bosons called
triple and quartic gauge couplings.
At LEP2 energies, triple gauge couplings are directly observed 
while quartic gauge couplings are negligible.
Interactions between neutral gauge bosons do not exist in the Standard Model.\\
\section{Triple Gauge Boson Couplings}
\noindent
The most general Lorentz invariant Lagrangian involving $WWV$ ($V=\gamma,Z)$ vertices can be
parametrized by 14 real parameters~\cite{Hagiwara}
\begin{equation}
i~\mathcal{L}^{WWV}/~g_{WWV}=
g_{1}^{V}V^{\mu}(W_{\mu \nu}^{-}W^{+\nu}-W_{\mu \nu}^{+}W^{- \nu})+
\kappa_{V} W_{\mu}^{+}W_{\nu}^{-}V^{\mu \nu}
\end{equation}
\begin{displaymath}
\hspace{0.5cm}
+~\frac{\lambda_{V}}{M_{W}^{2}}V^{\mu \nu}
W_{\nu}^{+ \rho}W_{\rho \mu}^{-} + 
ig_{5}^{V}\epsilon_{\mu \nu \rho \sigma}
[(\partial^{\rho} W^{- \mu})W^{+ \nu}-W^{- \mu}(\partial^{\rho} W^{+ \nu})]
V^{\sigma}
\end{displaymath}
\begin{displaymath}
\hspace{0.5cm}
+~ig_{4}^{V}W_{\mu}^{-}W_{\nu}^{+}(\partial^{\mu}V^{\nu}-\partial^{\nu}V^{\mu})
-\frac{\tilde{\kappa}_{V}}{2}W_{\mu}^{-}W_{\nu}^{+}
\epsilon^{\mu \nu \rho \sigma}V_{\rho \sigma}
-\frac{\tilde{\lambda}_{V}}{2M_{V}^{2}}\rm{W}^{-}_{\rho \mu}\rm{W}^{+ \mu}_{\nu}
\epsilon^{\nu \rho \alpha \beta} V_{\alpha \beta}
\end{displaymath}
where $g_{1}^{V}$, $\kappa_{V}$, $\lambda_{V}$ and $g_{5}^{V}$
are $\rm{CP}$-conserving couplings while
$g_{4}^{V}$, $\tilde{\kappa}_{V}$ and $\tilde{\lambda}_{V}$
are $\rm{CP}$-violating. Assumming $\rm{CP}$-conservation
and electromagnetic gauge invariance, five parameters are left : $g_{1}^{Z}$,
$\kappa_{Z}$, $\kappa_{\gamma}$, $\lambda_{Z}$ and 
$\lambda_{\gamma}$. The custodial $SU(2)$ symmetry  
of the Lagrangian imposes the constraints 
\begin{equation}
\hspace{-2.0cm}
\kappa_{Z}=g_{1}^{Z}-(\kappa_{\gamma}-1)\tan^{2}\theta_{W}
\hspace{2.0cm}
\lambda_{\gamma}=\lambda_{Z}
\end{equation}
where $\theta_{W}$ is
the weak mixing angle. Three free parameters are left : $g_{1}^{Z}$,
$\kappa_{\gamma}$ and $\lambda_{\gamma}$. They are 
related to the magnetic dipole and
electric quadrupole moment of the $\rm{W}$.
The Standard Model predicts their values to be
$g_{1}^{Z}=
\kappa_{\gamma}=1$ and $\lambda_{\gamma}=0$ at tree level.
\begin{figure}[t]
\begin{center}
%\rule{5cm}{0.2mm}
%\hfill\rule{5cm}{0.2mm}
%\vskip 2.5cm
%\rule{5cm}{0.2mm}\hfill\rule{5cm}{0.2mm}
%\hspace{-0.4in} 
\rotatebox{90}{\includegraphics[width=3.2cm,bb=448 776 168 16,clip]{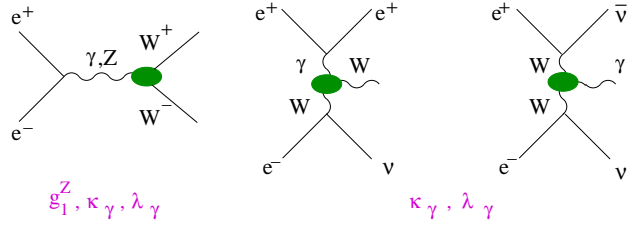}}
\caption{ The processes sensitive to $ZWW$ and $\gamma W W$ triple gauge couplings :
$\rm{W}$-pair, single $\rm{W}$ and single photon 
production.  }
\label{fig:tgcdia}
\end{center}
\end{figure}

\indent
The triple gauge couplings are studied in $\rm{W}$-pair production,
sensitive to all three parameters, and single $\rm{W}$ and single photon
production which are sensitive to $\kappa_{\gamma}$ and $\lambda_{\gamma}$ only.
The corresponding Feynman diagrams are presented in 
Figure~\ref{fig:tgcdia}. 
Only results from $\rm{W}$-pair and single $\rm{W}$
production are used at present for LEP combination.
\begin{figure}[b]
\begin{center}
%\rule{5cm}{0.2mm}
%\hfill\rule{5cm}{0.2mm}
%\vskip 2.5cm
%\rule{5cm}{0.2mm}\hfill\rule{5cm}{0.2mm}
%\hspace{-0.4in} 
\includegraphics[width=4.9cm,bb=0 95 609 696,clip]{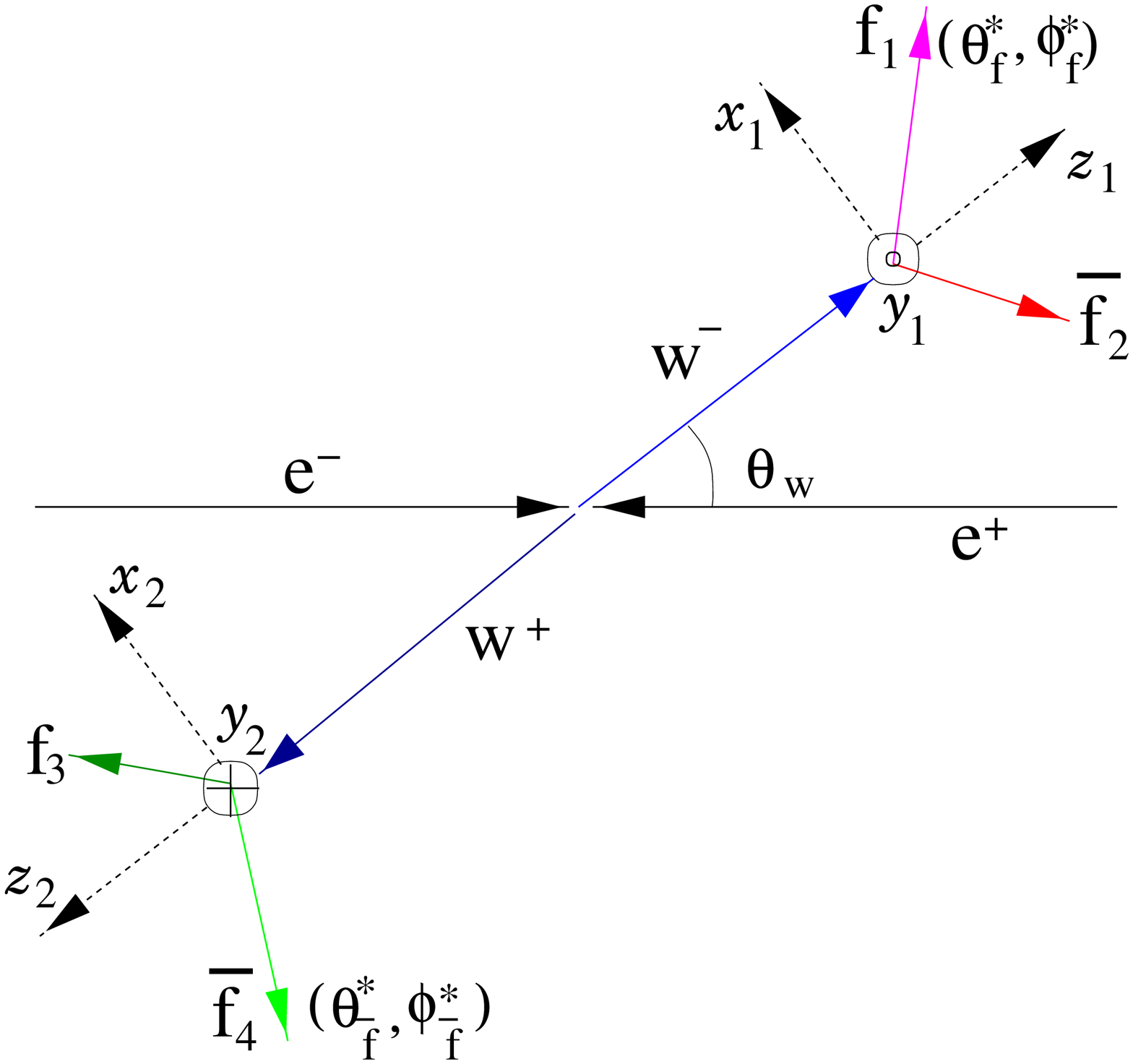}
\hspace{1.4in} 
\includegraphics[width=5.0cm,bb=40 128  570 670,clip]{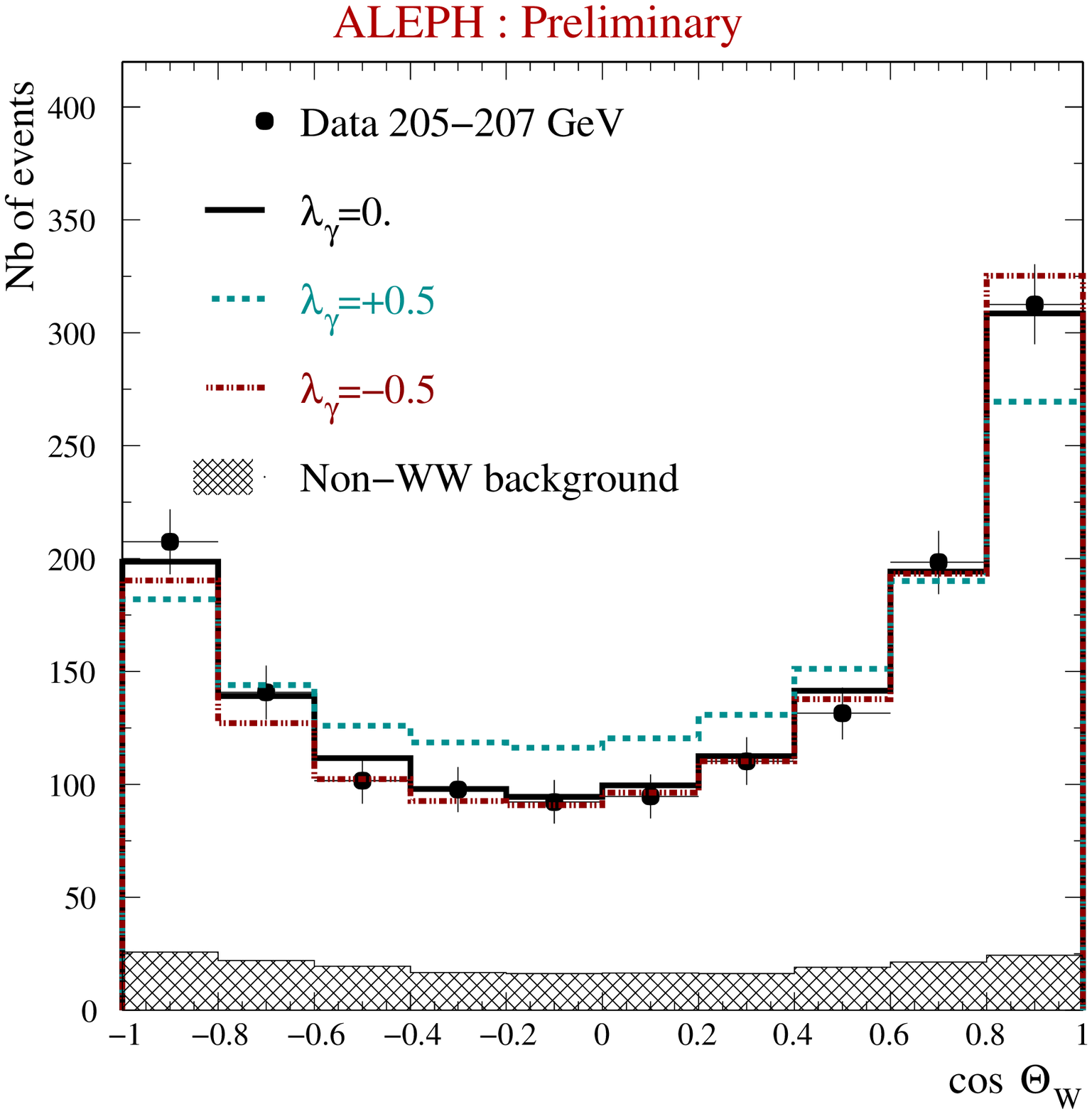}
\caption{ Left : The definition of the angles used in the coupling extraction 
Right : $\rm{W}$ production angle measured by the ALEPH experiment 
in fully hadronic $\rm{W}$-pair events.  }
\label{fig:sdmangles}
\end{center}
\end{figure}

\indent
A deviation from a coupling value predicted by the Standard Model 
would modify the total cross section, the shape of the $\rm{W}$
production angle $\theta_{W}$, the polar angle 
$\theta_{f}^{\star}$ and azimuthal angle,
$\phi_{f}^{\star}$ of the 
$\rm{W}$-decay fermions in the corresponding $\rm{W}$ rest frame.
These angles are presented in Figure~\ref{fig:sdmangles}
together with the distribution of the $\rm{W}$
production angle as measured by the ALEPH experiment 
in fully hadronic $\rm{W}$-pair events. The expected distribution
in presence of an anomalous coupling
$\lambda_{\gamma}=\pm 0.5$ is also indicated.

\indent
The couplings are extracted by a maximum likelihood fit 
to the angular distributions (DELPHI, L3)
or by a $\chi^{2}$-fit to Optimal Observables distributions
(ALEPH, OPAL). 
The results from each LEP experiment are then combined using a log-likelihood
method ~\cite{LEPEWWG,JUAN}.
\indent
\begin{figure}[t!]
\begin{center}
%\rule{5cm}{0.2mm}
%\hfill\rule{5cm}{0.2mm}
%\vskip 2.5cm
%\rule{5cm}{0.2mm}\hfill\rule{5cm}{0.2mm}
%\hspace{-0.4in} 
\mbox{\includegraphics[width=4.8cm,bb=43 69 587 690,clip]{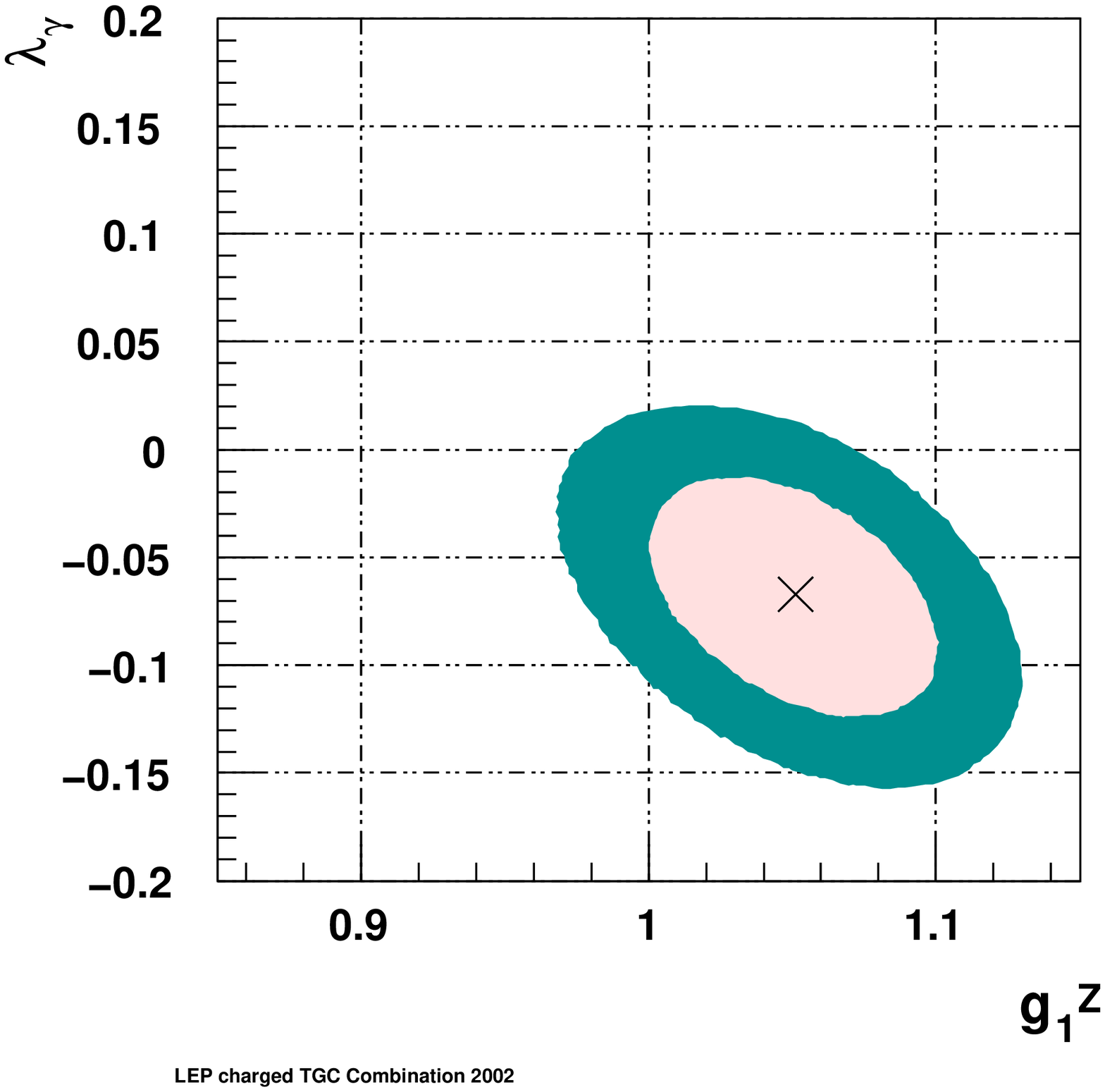}}
\mbox{\includegraphics[width=4.9cm,bb=19 75 563 714,clip]{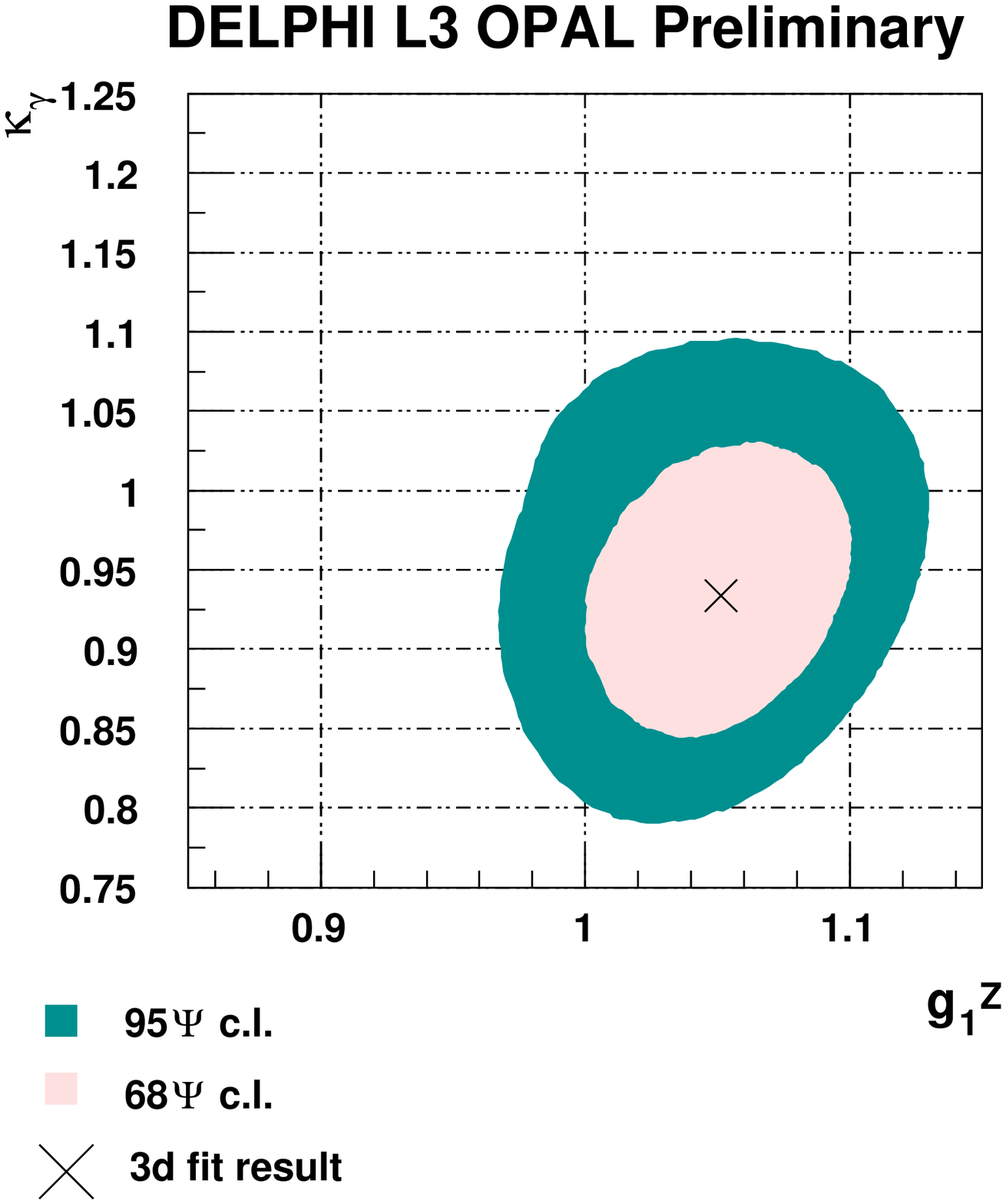}}
\mbox{\includegraphics[width=4.8cm,bb=30 49 585 672,clip]{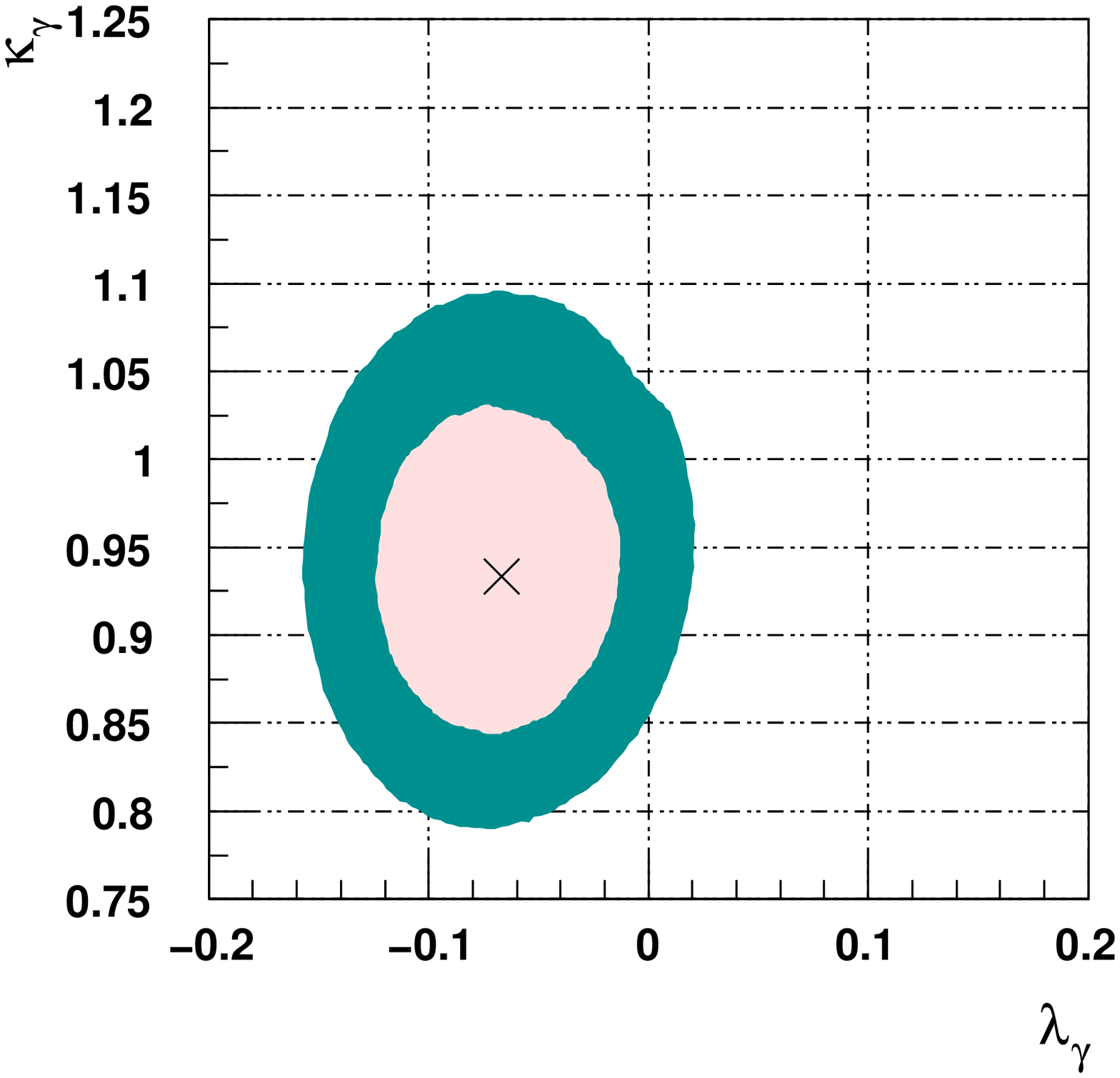}}
\caption{The result of the three-parameter fit, plotted
in two-dimensional planes defined by the three TGC pairs. The 
third parameter is at its fitted value.}
\label{fig:threeparfit}
\end{center}
\end{figure}

\indent
The result from the
three-parameter fit, including LEP2 data from DELPHI, L3 and OPAL, is presented in
Figure~\ref{fig:threeparfit}.
The LEP combined one-parameter fit results~\cite{LEPEWWG}
\begin{center}
$g_{1}^{Z}= 0.998^{+.023}_{-.025}$\hspace{0.75cm}
$\kappa_{\gamma}= 0.943^{+.055}_{-.055}$\hspace{0.75cm}
$\lambda_{\gamma}= -0.020^{+.024}_{-.024}$~.
\end{center}

\noindent
are in agreement with the Standard Model prediction.
The quoted errors include both statistical and systematic uncertainties.
In the LEP combination all $\rm{W}$ decay channels were used
except the semi-leptonic channel for L3 and LEP2 data from 
$\sqrt{s}=189$ GeV on for DELPHI.\\
\begin{table}
\caption{The sources of experiment correlated 
systematic uncertainties in the LEP combination and their effect on the combined fit
results.\label{tab:sys}}
\vspace{0.4cm}
\begin{center}
\begin{tabular}[ht!]{|c|c|c|c|}
\hline
\textbf{Source} & \textbf{$g_{1}^{Z}$} 
&\textbf{$\lambda_{\gamma}$} & 
\textbf{$\kappa_{\gamma}$}\\
\hline
$\mathcal{O}(\alpha)$ corrections & 0.015 & 0.015 & 0.039\\
Hadronisation & 0.004 & 0.002 & 0.004 \\
Bose-Einstein correlation & 0.005 & 0.004 & 0.009 \\
Colour Reconnection & 0.005 & 0.004 & 0.010 \\
$\sigma_{WW}$ prediction& 0.003 & 0.005 & 0.014 \\
$\sigma_{single~W}$ prediction&  & & 0.011 \\
\hline
\end{tabular}
\end{center}
\end{table}
\indent
The correlated 
systematic uncertainties in the combined fit
results are presented in Table~\ref{tab:sys}.
The largest contribution comes from the $\mathcal{O}(\alpha)$ radiative
corrections, mainly due
to virtual radiative corrections between the $\rm{W}$ bosons and the initial and final state
particles. 
The effect is taken into account by Monte Carlo generators
as YFSWW3~\cite{YFSWW3} in the Leading Pole Approximation (LPA)
and RacoonWW~\cite{RACOON}
in the Double Pole Approximation (DPA) and
introduces a 0.7 \% uncertainty on the slope of the $\cos\theta_{W}$ distribution.
Up to now the full difference between the Monte Carlo prediction
with and without $\mathcal{O}(\alpha)$ radiative corrections is taken as a systematic uncertainty.
Studies are still going on in the LEP experiments to define a more realistic estimate of the
$\mathcal{O}(\alpha)$ uncertainty.
An update on the TGC fits is planned for summer 2003.

\indent
\begin{table}
\caption{The LEP combined one-dimensional limits at 95 \% confidence level
for the neutral triple gauge couplings.\label{tab:hf}}
\vspace{0.4cm}
\begin{center}
\begin{tabular}[hb]{|c|c|c|}
\hline
\textbf{\ 95\% CL} & \textbf{\ $V=Z$}
& \textbf{\ $V=\gamma$} \\
\hline
\
$f_{4}^{V}$    &\  [ -.31 ; .28 ] & \ [ -.17 ; .19 ] \\
$f_{5}^{V}$    & \ [ -.36 ; .39 ] & \ [ -.36 ; .40 ]\\
\hline
\ $h_{1}^{V}$    & \ [ -.13 ; .13 ] & \ [ -.06 ; .06 ]\\
\ $h_{2}^{V}$    & \ [ -.08 ; .07 ] & \ [ -.05 ; .03 ]\\
\ $h_{3}^{V}$    & \ [ -.20 ; .07 ] & \ [ -.05 ; -.01 ] \\
\ $h_{4}^{V}$    & \ [ -.05 ; .12 ] & 
\ [ -.002 ; .034 ]\\
\hline
\end{tabular}
\end{center}
\end{table}

\indent
Neutral triple gauge couplings do not exist in the Standard Model.
The most general Lorentz invariant Lagrangian~\cite{Hagiwara,GOUR} for the
$VVZ$ ($V=\gamma,Z)$ vertex is described by 12 parameters. The couplings
$h_{1}^{V},~h_{2}^{V}$,
$h_{3}^{V}$ and $h_{4}^{V}$ are studied at LEP in the $\rm{e}^{+}\rm{e}^{-}
\rightarrow Z\gamma$
production, while  $f_{4}^{V}$ and $f_{5}^{V}$ are accessible
in $\rm{e}^{+}\rm{e}^{-}
\rightarrow ZZ$ production. Electromagnetic gauge invariance and
Bose symmetry for final states with identical bosons are imposed.
The couplings are determined from the angular distributions of the decay products and
the total cross section. No evidence for anomalous $h$- and $f$- couplings has been found.
The LEP combined one-dimensional limits at 95 \% confidence level are~\cite{LEPEWWG,JUAN}
summarized in Table~\ref{tab:hf}.
Both statistical and systematic uncertainties are included.

\section{{\boldmath $W$} Spin Density Matrix}
The Spin Density Matrix (SDM) method~\cite{Gournaris1} has been introduced
to study the $\rm{W}$ polarisation and is also used 
to set direct limits on 
$\rm{CP}$-violating couplings, absent in the Standard Model.  

\indent
Considering the helicity, the $\rm{W}$-pair production process is written as
\begin{equation}
\rm{e}^{+}(\lambda')~\rm{e}^{-}(\lambda) \rightarrow
\rm{W}^{+}(\tau_{2})~\rm{W}^{-}(\tau_{1})~,
\end{equation}
where $\lambda~(\lambda')=\pm 1/2$ 
represents the helicity of the electron (positron).
The helicities of the $\rm{W}^{-}$ and the $\rm{W}^{+}$, denoted by
$\tau_{1}$ and $\tau_{2}$ respectively,
take the value
$\tau=\pm 1$ for transversely polarised $\rm{W}$ bosons and the value $\tau=0$ for $\rm{W}$ bosons 
with a longitudinal polarisation.

\indent
The two-particle joint SDM elements are then defined as~\cite{Gournaris2,Bilenky}  
\begin{equation}
\rho_{{ \tau_{1}} { \tau_{1}}' { \tau_{2}} { \tau_{2}}'} (s,\cos \theta_{W})
\equiv
\frac{
\sum_{\lambda}
F^{\lambda}_{{ \tau_{1}} { \tau_{2}}}
~(F^{\lambda}_{{ \tau_{1}}' { \tau_{2}}'})^{\star}
}
{
\sum_{\lambda, { \tau_{1}}, { \tau_{2}}}
|F^{\lambda}_{{ \tau_{1}} { \tau_{2}}} |^{2}
}~,
\end{equation}
where $s$ is the center of mass energy and $F^{\lambda}_{\tau_{1} \tau_{2}}$ 
is the helicity amplitude for the production
of a $\rm{W}$ pair with helicities $\tau_{1}$ and $\tau_{2}$.
The single particle SDM elements are obtained
by summation over all possible helicities of one of the $\rm{W}$'s 
\begin{equation}
\rho_{{ \tau_{1}} { \tau_{1}}'}^{\rm{W}^{-}}(s,\cos
\theta_{\rm{W}^{-}})
\equiv
\sum_{{ \tau_{2}}}
\rho_{{ \tau_{1}} { \tau_{1}}'
{ \tau_{2}} { \tau_{2}}} (s,\cos \theta_{\rm{W}^{-}})~.
\end{equation}
The SDM elements are constrained by Hermiticity and 
their diagonal terms are normalised to unity 
\mbox{$\sum_{\tau} \rho_{ \tau \tau}^{\rm{W}^{-}}=1$}.
The diagonal elements of the SDM matrix are real and express the probability to produce
a $\rm{W}^{-}$ with helicity $\tau_{1}$.
The off-diagonal elements are complex and provide a test of
$\rm{CP}$-violation. 

\indent
The SDM elements are calculated in bins of 
$\cos\theta_{\rm{W}^{-}}$ using a projection operator method~\cite{Gournaris2}
assuming a V-A decay of the $\rm{W}$ boson into fermions
\begin{equation} 
\rho_{\tau \tau '}^{\rm{W}^{-}}
(k)
=\frac{1}{N_{k}}\sum_{i=1}^{N_{k}}
\Lambda^{\rm{W}^{-}}_{\tau \tau'}
(\theta^{\star}_{f},\phi^{\star}_{f})_{i}~,
\end{equation}
where $N_{k}$ is the number of events in the $k$-th bin and where
the projection operator $\Lambda^{\rm{W}^{-}}_{\tau \tau'}$
is applied event by event.
The reconstructed SDM elements need to be corrected for detector acceptance,
resolution effects and background contamination for
a direct comparison with the theoretical expectation.
\begin{figure}[t]
  \begin{center}
\includegraphics[width=7.99cm,bb=0 0 530 641,clip]{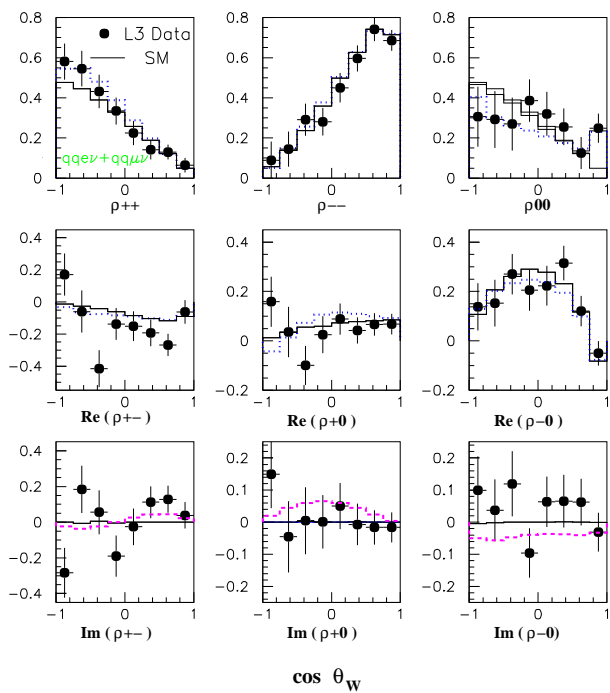}
  \end{center}
  \caption{The nine single SDM elements, $\rho_{\tau \tau'}^{\rm{W}^{-}}$, as a function of
$\cos\theta_{\rm{W}^{-}}$. 
The errors are statistical only.}
  \label{fig:sdmqqln}
\end{figure}

\indent
The SDM elements measured by L3~\cite{LSDM} in the decay channels
$q\bar{q}e\nu_{e}$ and the $q\bar{q}\mu\nu_{\mu}$ at 
$\sqrt{s}=189-209$ GeV are combined and 
presented in Figure~\ref{fig:sdmqqln}.
The measurements for the leptonically decaying $\rm{W}^{+}$ and $\rm{W}^{-}$ are combined assuming
$\rm{CPT}$-invariance. 
A good agreement is found with the Standard Model prediction represented by the solid line.
The expected distributions in presence of an anomalous $\rm{CP}$-conserving coupling $\Delta \kappa_{\gamma}=+0.5$ 
(dotted line) and the
$\rm{CP}$-violating coupling
$\tilde{\lambda}_{Z}=-0.5$ (dashed line) are also shown.\\
\begin{figure}[htbp]
  \begin{center}
    \includegraphics[width=7.8cm,bb=34 92 583 707,clip]{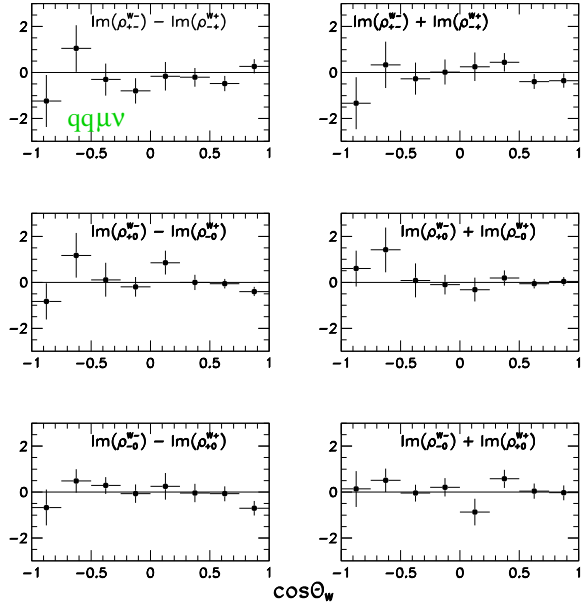}
  \end{center}
  \caption{ The sum of the imaginary parts of the off-diagonal
elements, sensitive to $\rm{CPT}$-violation at tree level (left),
and the difference, sensitive to $\rm{CP}$-violation at tree level (right).
The errors are statistical only.}
  \label{fig:violqqln}
\end{figure}
\indent
The imaginary parts of the off-diagonal elements
are insensitive to $\rm{CP}$-conserving couplings and
only contribute in presence of tree level $\rm{CP}$-violation.
This makes the SDM method particularly suitable to measure $\rm{CP}$-violating couplings  
which are extracted by a $\chi^{2}$-fit
to the nine SDM-element distributions. As the $\cos\theta_{W}$ information is averaged out in the
definition of the SDM elements, the shape of the $\rm{W}$ production angle is incorporated in the fit
to increase the sensitivity. 
The following results are obtained by the OPAL~\cite{OSDM} experiment
with the $q\bar{q}l\nu_{l}$ events selected at 189 GeV 
\begin{center}
$g_{4}^{Z}= -0.01^{+.32}_{-.33}$\hspace{0.75cm}
$\tilde{\kappa}_{\gamma}= -0.18^{+.24}_{-.16}$\hspace{0.75cm}
$\tilde{\lambda}_{\gamma}= -0.20^{+.10}_{-.07}$~.
\end{center}
All couplings are set to their Standard Model value except the measured one and these related
to it by custodial $SU(2)$ symmetry.
Both statistical and systematic uncertainties are included.

\indent
In the Standard Model $\rm{W}$-pair production is assumed to be a 
$\rm{CPT}$- and $\rm{CP}$- invariant
process, hence the following relations are satisfied\\
\begin{equation}
 {\rm CPT-invariance : \mathcal{I}m(\rho_{ \tau \tau'}^{\rm{W}^{-}})+
\mathcal{I}m(\rho_{ -\tau -\tau'}^{\rm{W}^{+}})=0}
\end{equation}

\vspace{-0.25in}
\begin{equation}
 {\rm CP-invariance : \mathcal{I}m(\rho_{ \tau \tau'}^{\rm{W}^{-}})-
\mathcal{I}m(\rho_{ -\tau -\tau'}^{\rm{W}^{+}})=0~}
\end{equation}
The imaginary part of all SDM elements has to be zero
and deviations from equation (8)
provide an unambiguous signature 
for $\rm{CP}$-violation at tree level.
Deviations from equation (7) would arise from loop effects beyond tree level 
or $\rm{CPT}$-violation.
Figure~\ref{fig:violqqln} shows the test of $\rm{CPT}$-invariance (right) and 
$\rm{CP}$-invariance (left)
measured with $q\bar{q}\mu\nu_{\mu}$ events selected by 
DELPHI~\cite{DSDM}  at  
$\sqrt{s}=189$ GeV.
Within the statistical error, the sum,  as well as 
the differences, of the imaginary parts 
are compatible with zero and confirm the absence of $\rm{CPT}$- 
and $\rm{CP}$-violation at
tree level as predicted by the Standard Model (solid line). This is confirmed by the 
L3~\cite{LSDM} and OPAL~\cite{OSDM}
results.

\indent
In DELPHI
and OPAL, the SDM analysis is also used to measure the cross sections for the
production of transversely and longitudinally polarised $\rm{W}$ bosons which are a consequence of 
the spontaneous symmetry breaking mechanism in the Standard Model.
The fraction of longitudinally polarised $\rm{W}$ bosons measured
by OPAL at 189 GeV in the $q\bar{q}l\nu$ channel
is $\sigma_{L}$/$\sigma_{total}= 21.0\pm3.3\pm1.6$ \%
where the first error is statistical and the second systematic.
This is in agreement with the Standard Model expectation of 25.7\%.

\indent
In L3~\cite{Radek}, the $\rm{W}$ polarisation is measured by a direct fit of analytical helicity distributions
to the shape of the polar angle $\theta^{\star}_{f}$ of the $\rm{W}$ decay products in semi-leptonic 
$\rm{W}$ pair events.
\begin{figure}[t]
  \begin{center}
\includegraphics[width=7.8cm,bb=0 0 530 641,clip]{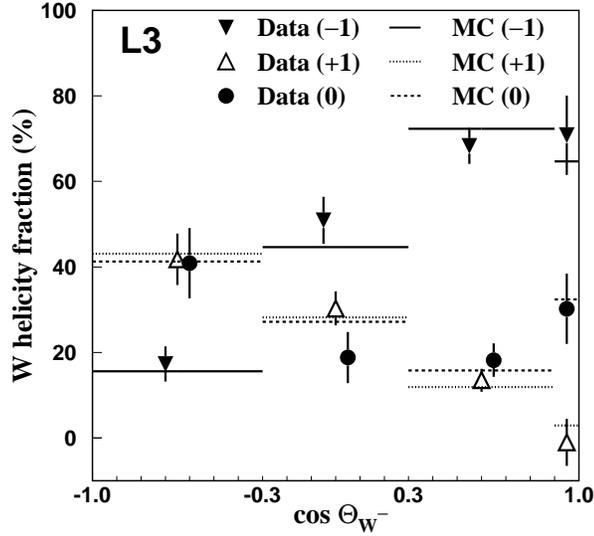}
  \end{center}
  \caption{The $\rm{W}$ helicity fractions in four different bins of 
$\cos\theta_{\rm{W}^{-}}$ measured with the L3 detector
using $q\bar{q}l\nu$ events at $\sqrt{s}=183-209$ GeV   
compared to the Standard Model predictions from the KORALW Monte $\rm{Carlo}^{15}$.}
  \label{fig:radek}
\end{figure}
The $\rm{W}$ helicity fractions are presented in Figure~\ref{fig:radek} in four different bins of 
$\cos\theta_{\rm{W}^{-}}$.
The fraction of longitudinally polarised $\rm{W}$ bosons
measured with the L3
detector using $q\bar{q}l\nu$ events at $\sqrt{s}=183-209$ GeV 
is $21.8\pm2.7\pm1.6$\% and in agreement with the Standard Model expectation of 24.1\%.
Separate analyses of the $\rm{W}^{+}$ and $\rm{W}^{-}$
events are consistent with $\rm{CP}$-conservation.
\section{Quartic Gauge Couplings}
The Standard Model quartic gauge couplings contribution are too small to
be seen at LEP and any deviation is therefore a hint for new physics.
Deviations are introduced
into the Lagrangian~\cite{BELANGER,STIRLING} as effective couplings
at a new physics scale $\Lambda$. 

\indent 
Starting from electromagnetic gauge invariance and 
custodial $SU(2)$ symmetry, the most general Lorentz invariant 
Lagrangian
has 5 parameters. The quartic gauge couplings
$a_{0}^{W}/\Lambda^{2}$, $a_{n}^{W}/\Lambda^{2}$
and $a_{c}^{W}/\Lambda^{2}$ are studied in the 
$\rm{e}^{+}\rm{e}^{-} \rightarrow \rm{W}^{+}\rm{W}^{-}\gamma$ process
and in  $\rm{W}$ fusion 
into a final state with two photons and missing energy due to the emission of two neutrino's, while  
the neutral quartic gauge couplings $a_{0}^{Z}/\Lambda^{2}$
and $a_{c}^{Z}/\Lambda^{2}$, not existent in the Standard Model,
are searched for in the 
$\rm{e}^{+}\rm{e}^{-} \rightarrow \rm{ZZ} \gamma$ process.
\begin{table}
\caption{The one-dimensional limits on quartic gauge couplings 
set by DELPHI, L3 and OPAL at 95 \% CL.\label{tab:qgc}}
\vspace{0.4cm}
\begin{center}
\begin{tabular}[b!]{|c|c|c|c|}
\hline
95 \% CL ( ${\rm GeV}^{-2}$) &
 $ a_{0}^{W}/\Lambda^{2}$ & 
$a_{c}^{W}/\Lambda^{2}$ &  $a_{n}^{W}/\Lambda^{2}$ \\
\hline
DELPHI &  [ -.018 ; +.018] & 
 [ -.057 ; +.030 ] & [ -.16 ; +.12 ] \\
 L3  &
 [ -.015 ; +.015 ]   &  [ -.048 ; +.026 ]
 & [ -.14 ; +.13 ] \\
 OPAL 
 &  [ -.054 ; +.052 ] &
  [ -.15 ; +.14 ]  & 
 [ -.61; +.57 ] \\
\hline
\end{tabular}
\end{center}
\end{table}
Both are mainly determined from the photon energy spectrum and
the total cross section.
 
\indent
No evidence for anomalous quartic gauge couplings has been found.
The one-dimensional limits on quartic gauge couplings
set by DELPHI~\cite{DQGC}, L3~\cite{LQGC} 
and OPAL~\cite{OQGC}
at 95 \% confidence level are 
summarized in Table~\ref{tab:qgc}.
Both statistical and systematic uncertainties are included.
The one-dimensional limits on neutral quartic gauge couplings set by L3 and OPAL 
at 95 \% confidence level are~\cite{LEPEWWG}
\begin{displaymath}
-.009~<a_{0}^{Z}/\Lambda^{2}~.~{\rm GeV^{2}}<~.026
~~-.033~<a_{c}^{Z}/\Lambda^{2}~.~{\rm GeV^{2}}< ~.046
\end{displaymath}
New results from ALEPH~\cite{AQGC} yield
\begin{displaymath}
-.011~<a_{0}^{Z}/\Lambda^{2}~.~{\rm GeV^{2}}<~.017
~~-.037~<a_{c}^{Z}/\Lambda^{2}~.~{\rm GeV^{2}}< ~.040
\end{displaymath}
at 95 \% confidence. 
Both statistical and systematic uncertainties are included.
A LEP combination of the quartic gauge couplings is expected soon.

\section*{Acknowledgments}
I wish to express my gratitude to my collegues from LEP and convenors of the LEP
gauge coupling working groups for their support and discussions.

\end{document}